\documentclass[12pt]{article}
\usepackage{lipsum}
\usepackage{refstyle}
\usepackage{amsfonts}
\usepackage{amsthm}
\usepackage{amsmath}
\usepackage{mathtools}
\usepackage{graphicx}
\usepackage{setspace}
\usepackage{soul}
\usepackage[authoryear]{natbib}
\usepackage[margin=1in]{geometry}
\usepackage{color}
\usepackage{bbm}
\usepackage{booktabs}
\usepackage{pdflscape}
\usepackage{dcolumn}
\usepackage{threeparttable}
\usepackage{siunitx}
\usepackage{caption}
\usepackage{floatrow}
\usepackage{array}
\usepackage{appendix}
\usepackage{enumitem}
\usepackage{subfig}
\usepackage[countmax]{subfloat}
\usepackage{placeins}
\usepackage{fancyhdr}
\usepackage[breaklinks=true]{hyperref}
\usepackage{cases}
\usepackage{etoolbox}
\usepackage{sectsty}
\usepackage{titlesec}
\usepackage{ragged2e}
\usepackage{lmodern,textcomp}
\usepackage{tikz,pgfplots}
\usetikzlibrary{arrows,automata,positioning}
\pgfplotsset{compat=1.16}
\usetikzlibrary{calc,patterns,angles,quotes}

\sectionfont{\normalfont\scshape\centering}
\subsectionfont{\normalfont}

\titlespacing*{\section}{0pt}{2.5ex}{2.5ex}
\titlespacing*{\subsection}{0pt}{2ex}{2ex}
\titlespacing*{\subsubsection}{0pt}{2ex}{2ex}

\makeatletter
\renewcommand*{\@seccntformat}[1]{\csname the#1\endcsname\hspace{0.5em}}
\makeatother

\definecolor{darkblue}{rgb}{0,0,0.5}
\definecolor{darkred}{rgb}{0.7,0,0}
\hypersetup{
	colorlinks,
	citecolor=darkblue,
	linkcolor=darkred,
	urlcolor=darkblue
}

\def\equationautorefname~#1\null{(#1)}

\makeatletter
\renewcommand*{\@fnsymbol}[1]{\ensuremath{\ifcase#1\or *\or \dagger\or \ddagger\or
   \mathsection\or \mathparagraph\or \|\or **\or \dagger\dagger
   \or \ddagger\ddagger \else\@ctrerr\fi}}
\makeatother

\makeatletter
\g@addto@macro\normalsize{%
  \setlength\belowdisplayshortskip{\belowdisplayskip}
}
\makeatother

\makeatletter
\patchcmd{\hyper@makecurrent}{%
    \ifx\Hy@param\Hy@chapterstring
        \let\Hy@param\Hy@chapapp
    \fi
}{%
    \iftoggle{inappendix}{
        \@checkappendixparam{chapter}%
        \@checkappendixparam{section}%
        \@checkappendixparam{subsection}%
        \@checkappendixparam{subsubsection}%
        \@checkappendixparam{paragraph}%
        \@checkappendixparam{subparagraph}%
    }{}%
}{}{\errmessage{failed to patch}}

\newcommand*{\@checkappendixparam}[1]{%
    \def\@checkappendixparamtmp{#1}%
    \ifx\Hy@param\@checkappendixparamtmp
        \let\Hy@param\Hy@appendixstring
    \fi
}
\makeatletter

\newtoggle{inappendix}
\togglefalse{inappendix}

\apptocmd{\appendix}{\toggletrue{inappendix}}{}{\errmessage{failed to patch}}
\apptocmd{\subappendices}{\toggletrue{inappendix}}{}{\errmessage{failed to patch}}

\renewcommand{\thefigure}{\arabic{figure}}

\DeclareCaptionLabelFormat{panel_label}{Panel #2.}
\usepackage{multirow}
\newcommand{\Tabnote}[1]{
	\vspace*{-0.15cm}
	\begin{tablenotes}[para, flushleft]
		\footnotesize\textsc{Note.---}{#1 \hfill}
	\end{tablenotes}	
}

\newcolumntype{L}[1]{>{\raggedright\let\newline\\\arraybackslash\hspace{0pt}}m{#1}}
\newcolumntype{C}[1]{>{\centering\let\newline\\\arraybackslash\hspace{0pt}}m{#1}}
\newcolumntype{R}[1]{>{\raggedleft\let\newline\\\arraybackslash\hspace{0pt}}m{#1}}

\begin{document}

	% title page
	\begin{titlepage}
\renewcommand{\baselinestretch}{1}
    \newcommand\blfootnote[1]{%
  \begingroup
  \renewcommand\thefootnote{}\footnote{#1}%
  \addtocounter{footnote}{-1}%
  \endgroup
}
    % title
    \title{
        \sc{
            \Large{
            Insurance Supervision under Climate Change:\\ A Pioneer Detection Method
            }
        }
    }
    
    \author{
        \textbf{Eric Vansteenberghe}%
        \thanks{
            Banque de France: %
            \href{mailto:eric.vansteenberghe@banque-france.fr}%
            {\nolinkurl{eric.vansteenberghe@banque-france.fr}}
        }
    }
    
    % date
    \date{}
    
    \maketitle

        \blfootnote{
        We thank Antoine Baena, Paul Beaumont, Yvan B\'{e}card, Catherine Bobtcheff, Arthur Charpentier, Jean-Bernard Chatelain, Laurent Clerc, Roger Cooke, Jeffrey Czajkowski, Franz Dietrich, Christian Gourieroux, Olena Havrylchyk, Carolyn Kousky, Mirko Kraft, Jean-Gabriel Lauzier, Antonin Mac\'{e}, Nicolas Pasquier, Cyril Pouvelle, Pierre-Charles Pradier, Caroline Rufin-Soler, Andr\'{e} Schmitt, Isma\"{e}l Tahri-Hassani, Huan Tang, and Aluna Wang for their time and advice. We also thank participants at the 2022 ACPR scientific committee, the 2022 French Treasury conference on public policy evaluation, the 2023 Welfare and Policy Conference in Bordeaux, the July 2023 research seminar at the ACPR, the 2023 ARIA conference, the 2023 conference on the Economics of Climate Change and Environmental Policy, the 2023 Climate Finance, Risk, and Uncertainty Modeling conference, and the 2024 EGRIE conference. This work has benefited from the support of the Agence Nationale de la Recherche through the program Investissements d'Avenir ANR-17-EURE-0001.
    }
    
    \begin{abstract}
    \noindent
    We present the Pioneer Detection Method, a supervisory tool we developed to enhance resilience in insurance markets facing the challenges posed by climate change. Based on a theoretical model of the insurance industry, we consider a scenario in which independent experts determine premiums according to their individual risk assessments. Due to the segmented nature of the private insurance market, accurately estimating the tail parameter of loss distribution is difficult, especially given the rarity of extreme events. Our method leverages temporal directional change and convergence to integrate expert opinions, giving greater emphasis to those who effectively identify trend shifts after climate stress. A series of simulations reveals that the Pioneer Detection Method outperforms traditional pooling methods within a Bayesian framework. Furthermore, this approach appears to be notably effective in improving welfare in an insurance market with a limited number of private entities.
    %We introduce the Pioneer Detection Method, a framework designed to strengthen the resilience of insurance markets exposed to climate-related risks. Building on a micro-theoretical model of heterogeneous insurers, the method considers an environment in which independent experts set risk-based premiums according to their individual assessments of low-frequency, high-severity losses. Because private insurance markets are segmented and extreme events are rare, estimating the tail parameter of the loss distribution is subject to substantial model uncertainty and disagreement. The Pioneer Detection Method aggregates expert beliefs using directional change, temporal convergence, and trend-shift detection after climate stress, assigning greater weight to experts who identify structural breaks or upward drifts in climate risk. In a series of simulations calibrated to climate-driven loss processes, the method consistently outperforms standard Bayesian pooling and conventional averaging rules. It improves risk measurement, premium adequacy, and market-wide welfare, particularly in markets with a small number of insurers where learning frictions and belief heterogeneity are most acute. The framework offers a practical supervisory tool for climate-change adaptation, early identification of emerging risks, and enhanced prudential oversight under Solvency II–type regimes.

    \bigskip
    \noindent
    \textbf{Keywords:} Climate change, insurance market stability, opinion pooling. 
    
    \noindent
    \textbf{JEL codes:} G22, G28, Q54.
    \end{abstract}
    
    \thispagestyle{empty}
\end{titlepage}

	% sections
	\section*{Extended Abstract}

\subsection*{1. Research Problem}
Climate change is altering the underlying loss-generating processes faced by insurers, leading to shifts in the distribution of yearly aggregated losses reported across firms. Because private insurance markets are segmented, each insurer observes only its own limited claims history, making it difficult to infer the tail parameter governing extreme losses. Climate-driven structural changes—whether gradual trends or sudden tipping points—introduce additional uncertainty, which different experts internalize at different speeds. When reinsurance capacity retracts after climate shocks, supervisors need tools that enable rapid learning, aggregation of heterogeneous beliefs, and timely assessment of the insurability of affected asset classes.

\subsection*{2. Objective}
This paper introduces the Pioneer Detection Method (PDM), a supervisory opinion-pooling framework designed to:
\begin{itemize}
    \item identify experts who detect climate-driven structural changes earlier than others;
    \item accelerate collective learning about extreme-loss tail parameters; and
    \item support supervisory decisions on insurability, premium adequacy, and market stability under Solvency~II--type regimes.
\end{itemize}

\subsection*{3. Theoretical Framework}

\paragraph{Insurance Market Model.}
Annual losses follow a Pareto distribution with an unobservable and climate-sensitive tail parameter~$\alpha_t$. Independent insurance experts estimate~$\alpha_t$ from small, proprietary samples, generating heterogeneous beliefs. A climate tipping point induces a discrete shift in the true tail parameter. Supervisors cannot observe the true parameter and therefore must rely on expert signals.

\paragraph{Key Frictions.}
\begin{itemize}
    \item Fat-tailed loss distributions: slow statistical learning.
    \item Fragmented datasets: insurers do not share claims histories.
    \item Heterogeneous Bayesian updating: sample-dependent learning speeds.
    \item Reinsurance withdrawal: information pooling disappears when most needed.
\end{itemize}

\subsection*{4. Methodology: The Pioneer Detection Method (PDM)}
The PDM identifies \emph{pioneers}---experts whose estimates deviate from the consensus but toward whom other estimates converge over time.

\paragraph{Core Components.}
\begin{itemize}
    \item \textbf{Distance Reduction Test}: checks whether cross-expert differences shrink between periods.
    \item \textbf{Orientation Change Test}: detects whether changes in other experts’ estimates move toward expert $i$.
    \item \textbf{Attribution Ratio}: quantifies how much of the convergence is driven by movements of peers, not the pioneer.
\end{itemize}

\paragraph{Weighting Scheme.}
The supervisor’s pooled estimate is
\[
\hat{\alpha}_S^{\,t} = \sum_i w_i^{\,t} \, \hat{\alpha}_i^{\,t},
\]
with weights derived from convergence signals rather than forecast errors, as the true tail parameter is unobservable.

\subsection*{5. Computational Experiments}

\paragraph{Design.}
Monte Carlo simulations with Bayesian experts learning~$\alpha$ from Pareto losses. Scenarios vary:
\begin{itemize}
    \item the number of insurers $m$,
    \item the magnitude of the climate-induced shift in $\alpha$,
    \item the post-tipping steady state.
\end{itemize}

\paragraph{Benchmarks.}
The PDM is compared against linear opinion pooling, median forecasts, minimum-risk expert, Bayesian averaging, and alternative novel pooling rules.

\paragraph{Performance Metrics.}
Root Mean Square Error (RMSE) relative to the true (simulated) tail parameter~$\alpha$.

\paragraph{Findings.}
\begin{itemize}
    \item \textbf{Fastest post-shock convergence}: PDM identifies early learners and outperforms all pooling methods from the first usable period.
    \item \textbf{Superior performance in fat-tailed environments}: classical pooling fails precisely when extreme events matter most.
    \item \textbf{Stable accuracy} across tail parameters and market sizes.
    \item \textbf{Economic relevance}: welfare improvements (via reduced variance of $\hat{\alpha}_S$) are substantial in markets with $\leq 5$ insurers, where limited sample sizes make expert aggregation most valuable.
\end{itemize}

\subsection*{6. Policy Implications}
Supervisors can use the PDM to:
\begin{itemize}
    \item assess insurability after climate shocks;
    \item guide emergency regulatory responses;
    \item monitor emerging climate risks in real time;
    \item reduce uncertainty-driven premium spikes or coverage withdrawals.
\end{itemize}
The PDM provides a lightweight alternative when granular data collection is costly or infeasible, and it is suited for markets lacking reinsurance or facing sudden tail-risk inflation.

\subsection*{7. Contributions}
This paper contributes to:
\begin{itemize}
    \item insurance supervision under climate change;
    \item opinion pooling and combination forecasting;
    \item Extreme Value Theory in loss modeling;
    \item pioneership versus leadership in risk assessment.
\end{itemize}
The PDM offers a novel, implementable approach to synthesizing expert knowledge in environments with structural breaks and extreme events.

\subsection*{8. Keywords}
Climate change; insurance supervision; extreme value theory; Pareto tail index; opinion pooling; Bayesian learning; structural breaks; climate tipping points; tail risks; insurability; Solvency~II; prudential regulation; reinsurance withdrawal; heterogeneous beliefs; supervisory tools; welfare analysis.

\section{Introduction}

There have recently been many press articles raising blunt questions about the insurability of our planet and academic articles warning of imminent climate tipping point \citep{ditlevsen2023}. Climate change, characterized by long-term shifts in temperatures and weather patterns, leads to both direct (e.g., heatwaves, droughts) and indirect (e.g., desert expansion, wildfires, storms) extreme events with increasing frequency and intensity \citep{stott2016attribution}. There remains debate about whether climate change will result in continuous trends or unexpected shocks that challenge financial systemic stability \citep{Svartzman:2021up}. Financial supervisors face the challenge of understanding insurability risks \citep{Berliner:1985uz} in this volatile environment, making the role of pooled expert opinions particularly valuable. Pooling methods, as introduced by \citet{stone1961}, offer a way to combine expert judgments into a unified perspective. However, these approaches often struggle to address extreme event risks or to estimate parameters of highly uncertain loss distributions.Consequently, traditional pooling approaches face two major limitations: evolving inconsistencies in expert views and delayed decision-making regarding insurability. This paper proposes a novel opinion pooling approach tailored to these challenges, referred to as the Pioneer Detection Method. It models an insurance market to validate and estimate the welfare benefits of this approach.

We start by designing an insurance market model derived from \citet{raviv1992design}. We depart from it by introducing beliefs heterogeneity among insurance experts and study its equilibrium to determine the optimal insurance contracts when reinsurance is not available. When this heterogeneity arises from information limited to small private samples, we study remedies that an insurance supervisor can implement to limit potential welfare loss. In a segmented insurance market with no sharing of proprietary claim histories, each expert bases their estimation on a restricted private sample.  In a segmented market where claim histories are not shared, experts estimate risks from limited private samples. This slows learning of the true tail parameter, because extreme events are rare, but Bayesian updating ensures eventual convergence, so welfare losses arise from delayed learning rather than biased perception. Climate change impacts both the frequencies and magnitudes of events. This paper models non-cooperative private insurance experts with proprietary information sets and modeling expertise.\footnote{The European Court of Justice established in 1987 (Judgement of 27.1.198) that EU competition law is fully applicable to the insurance sector. Some exemptions were allowed by the Insurance BER'' (Regulation (EEC) No. 3932/1992) as Collaboration between insurance undertakings [$\ldots$] in the compilation of information (which may also involve some statistical calculations) allowing the calculation of the average cost of covering a specified risk [$\ldots$] makes it possible to improve the knowledge of risks [$\ldots$]. This can in turn facilitate market entry and thus benefit consumers.'' The ``Insurance BER'' was stopped on April 1, 2017. Automatic exemptions were discarded to avoid misapplications, considering that guidelines were sufficient \citep{stancke2017old}.} We model a tipping point \citep{gladwell2006tipping,lenton2008tipping} where reinsurers, exogenous to the market, decide to withdraw. After this distinct and unanticipated shock, there is a situation of heterogeneous beliefs among insurance experts as they update their risk models to determine whether, and under what conditions, they provide insurance coverage for the affected asset class. An insurance supervisor, mandated with ensuring financial stability, is also included in the model. The supervisor decides if there is a need to intervene in the insurance market. The goal of this intervention would be to limit welfare losses by influencing the terms under which insurance is provided for specific asset classes. To do so and for the supervisor to estimate the risk after a tipping point, the context calls for a new insurance supervision tool.

In this paper, we introduce an innovative supervisory tool, namely the Pioneer Detection Method (PDM). It is particularly designed for the constantly changing climate, where the absence of solid, empirical data challenges expertise evaluations grounded on past estimations. The primary objective of this method is to assign substantial weightage to dependable pioneers---experts who have consistently underscored the intensifying gravity of climate-centric risks. These experts employ transparent models and datasets, and their viewpoints receive tacit validation from their professional peers over time. The opinion pooling weights can be understood as unspoken inter-temporal votes shared among experts. The method's core strength lies in its dependency on directional shifts to configure weights, rather than a primary emphasis on reducing disparities between opinions.

In the model, yearly aggregated insurance losses for an asset class follow a Pareto distribution with an unobservable tail index and a normalized threshold. This simplifies the problem to a single tail parameter that is impacted by climate change. With this single parameter, it is possible to model an increase in the average expected loss and the extreme yearly losses (right tail of the distribution): $E_{\alpha^t}(X) = \frac{\alpha^t}{\alpha^t-1}$ and $Pr(X>x)= \left(\frac{1}{x} \right)^{\alpha^t}$. For the validation of the PDM, we introduce a unique and unexpected change in the tail parameter, caused by climate change. There was a past tail parameter $\alpha^-$ that had been well learned by all experts, at $t=0$, the tail parameter jumps to a value $\alpha$ closer to unity and then for the rest of the validation period stays at this value. There is an initial uncertainty about the new value of the tail parameter to learn, but for the full validation period, this value is stable, albeit unknown, and hence each expert receives observations from independent and identical extreme value distributions.

When the insurability of an asset class suddenly becomes uncertain \citep{knight1921risk}, and reinsurance is unavailable, two key challenges arise from a financial stability perspective. First, some insurance companies may underestimate the implications of the change, leaving themselves vulnerable to insolvency. This paper, however, focuses on a second, more systemic concern: non-cooperative behavior among insurers leading to sharp increases in premiums.
This second challenge can manifest in two ways. In the first, premiums may exceed the maximum willingness or financial capacity of insurance buyers, effectively making coverage unattainable. In the second, premiums approach the insurance coverage limit itself; as \citet{winter1994dynamics} notes, insurance supply remains positive at some price, but premiums in certain cases, such as asbestos removal, have been observed to reach as high as 90\% of the coverage limit. To determine whether the reduction in insurance supply is justified by the increase in risks or the inability of insurance experts to accurately estimate risks, the supervisor can decide to collect expertise using ad-hoc on-site inspections of insurance companies and pool expertise. Although there is a rich literature on stress tests \citep{Battiston:2017vj} and their use in learning about the insurance market and guiding supervisory actions, little research has been conducted on recommending ad-hoc inspections for pooling opinions for the same purpose.\footnote{The closest strategy to inspection is NAIC's Climate Risk Disclosure Survey, a voluntary risk management tool through which state regulators can request annual, non-confidential disclosure of insurers' assessment and management of their climate-related risks. Source: Moody's report on Insurance Conference 2022.} This paper aims to fill this gap by demonstrating how supervisors can use their regular supervision or ad-hoc inspection activities to reinforce their knowledge and understanding of climate change's disruptive potential for the insurance markets. %An additional application would be to apply the method on heterogeneous bottom-up stress test projections.\footnote{A bottom-up stress test involves individual insurance companies projecting their exposure under a predefined stressed scenario, which is uniformly applied across all participating entities. This approach allows for the assessment of each company's resilience to specific stress factors, considering their unique risk profiles and management strategies. Notably, the ACPR initiated its second climate stress test for the insurance sector on July 6, 2023. This exercise underscores the regulatory focus on understanding the climate-related financial risks within the sector. By leveraging the PDM, a supervisor could aggregate the varied projections of the tail parameter changes provided by insurers, thus forming a more cohesive and informed perspective on potential sector-wide impacts.}

More broadly, this paper contributes to the continuing discourse on insurance supervision in the context of climate change. In countries like France, debates exist on whether housing insurance state guarantees for clay soil risks should be scrapped and how a supervisor can act once the market is fully private. Currently, 40\% of the state natural catastrophe scheme (CatNat) is used to cover losses for houses exposed to the risk of Withdrawal-Swelling of Clay Soils (WSCS) under the effect of drought. The French Court of Auditors is advising the exclusion of WSCS from the CatNat regime, given that the unpredictable nature of the risk is no longer being addressed due to the effects of climate change \citep{cc2022}. If WSCS was removed from the CatNat regime and if external reinsurers decided to discontinue the coverage for certain areas, the supervisor could use the PDM to learn as fast as possible from insurance companies internal expertise. An alternative approach would be to use historical data from the scheme and build his own granular model. When granular data are not readily available, and the supervisor's capacity is limited in modeling the risk for each asset class, the novel methodology enhances supervision. It aggregates expertise at the insurance group level, rather than initiating procedures to obtain granular underwriting data for evaluating climate-related financial risks post-shock. This paper addresses the challenge of estimating losses from natural disasters, focusing on the unobservable tail parameter in yearly loss aggregates, which are private to each expert. The difficulties in obtaining accurate loss estimates are elucidated, highlighting challenges such as the lack of geographic exposure information, limitations in climate stress test data, and issues with asset holdings data.\footnote{Solvency II reporting, which provides data only by line of business, poses limitations on the identification of areas most affected by natural disasters and subsequent loss estimation. The ACPR has initiated a climate stress test for insurers, with templates available \href{https://acpr.banque-france.fr/sites/default/files/media/2023/10/13/tableau_exercice_assurances_acpr_2023_1110.xlsx}{here}. Despite offering some insights into geographic exposures to natural catastrophes, the voluntary nature of insurer participation and potential data use restrictions beyond the stress test's scope significantly limit its utility for external research purposes and supervisor estimation of the tail parameter. Additionally, open-source tools like CLIMADA from the EIOPA (\href{https://www.eiopa.europa.eu/tools-and-data/open-source-tools-modelling-and-management-climate-change-risks_en}{link}) and open data from Météo France (\href{https://meteo.data.gouv.fr}{here}) provide alternative data sources. However, challenges persist in accurately modeling loss impacts, with the potential for significantly overestimated loss projections compared to real losses. The lack of publicly available historical loss data by municipalities from the Central Reinsurance Company (CCR) further limits researchers' capacity to estimate actual losses. While combining the Emergency Events Database (EMDat) (\href{https://www.emdat.be}{link}) with Catastrophic Natural Events (CatNat) decrees could offer insights into specific event impacts, the absence of loss quantification in CatNat decrees and potential limitations in event coverage and heterogeneity may hinder the derivation of statistically significant econometric results.} While the United States Department of the Treasury's Federal Insurance Office (FIO) has considered collecting granular data from property and casualty insurers regarding homeowners' insurance, it has faced resistance. This work can also be applied to climate change direct and indirect effects on human health \citep{healthcc2024}.  A policy recommendation from this paper is to allow the use of this new tool to help supervisors determine if an asset class is insurable. Upon determining the insurability of an asset class, the supervisor can then advise the regulator on whether to utilize public or private insurance mechanisms, such as syndication or reinsurance, for coverage.

\textbf{Literature} The main contribution of this paper is on the leadership and pioneership literature. The etymology and historical application of the terms `leadership' and `pioneership' have origins rooted deeply in military contexts. According to the Oxford English Dictionary (2015), a `leader' was initially defined as an individual who `conducts, precedes as a guide', while a `pioneer' was described as `a member of an infantry group [...] ahead of an army [...] to clear terrain in readiness for the main body of troops'. In financial contexts, leadership connotes certain interactions, with some investors being influenced by the actions of their counterparts. This effect is so significant that it's considered a `first-order effect' \citep{devenow1996rational}. Such interactions might result in `information cascade', `when it is optimal for an individual, having observed the actions of those ahead of him, to follow the behavior of the preceding individual without regard to his own information' \citep{bikhchandani1992theory}. \citet{hirshleifer2003herd} categorises information cascade as a subset of herding, which is recognized as behavioural uniformity stemming from individual interactions. Notably, in financial literature, the concept of pioneership is largely overlooked. It emerges `where individuals act similarly due to a parallel, independent influence of a shared external factor' \citep{hirshleifer2003herd}. In relation to climate change, a distinction emerges between leaders and pioneers. Specifically, a `leader state has the explicit ambition to lead others, while a pioneer state’s priority is to develop its own pioneering activities without paying (much) attention to attracting followers.' \citep{liefferink2017}. In this paper, the supervisor assumes a leadership role, whether on a temporary or permanent basis, guiding insurers to address climate change challenges. Conversely, a pioneer is delineated as an expert who temporarily serves as an indicator (predictor) for his peers in the subsequent time step. The primary motive driving pioneership is the intent to assess risks judiciously and calibrate premiums to prevent financial downturns for the insurance entity. With our innovative Pioneer Detection Method (PDM), multiple experts might qualify as pioneers during a particular timeframe, albeit with varying magnitudes. It is worth noting that in our definition, pioneership lacks inertia; an expert's status as a pioneer at a given moment doesn't guarantee their continued pioneership in future periods. Our approach significantly diverges from conventional financial literature in two pivotal ways. First, experts, in our context, do not leverage information from fellow experts' premiums. As experts are in competition, their primary focus is on pioneership to bolster their insurance firm's fiscal health, rather than leadership or mimicking another expert. Only the insurance supervisor is equipped to gather information and discern the pioneership intensity of each expert. Second, unlike traditional financial paradigms, our approach acknowledges that the supervisor cannot ever truly discern the risk's actual realization. Hence, pioneership can only be discerned when an expert's actions seem to be emulated by peers over successive periods. Implicit in this is the belief that experts' precision will refine over time. An alternative perspective on pioneership intensity involves weighting experts to expedite convergence towards a more precise risk assessment.

This paper contributes to the literature on opinion pooling and combination forecasting, building upon previous reviews by \citet{genest1986}, \citet{clemen1989combining}, \citet{timmermann2006forecast}, and \citet{wang2022forecast}. This literature concludes that in a Gaussian context it is almost impossible to beat a mean method potentially excluding outliers. In a context with fat tails, extreme events do occur but are rare, hence learning is slow and we design a new method that relies on inter-temporal opinion changes to weigh experts that are the most likely to learn faster about extreme events.

This paper adds to the rich literature and ongoing debate surrounding the impacts of climate change on insurance losses. There are mixed conclusions on the effects of climate change on loss trends (especially after normalizing with the local gross domestic products), as discussed by \citet{mills2005insurance}, \citet{pielke2008normalized}, \citet{stern2008economics}, \citet{kousky2014informing}, \citet{hoeppe2016trends} \citet{hsiang2016climate}, and \citet{botzen2020economic}. This unresolve academic debate is influencing climate change perception and action, innovative approaches like climate prediction markets have been shown to significantly alter public attitudes towards climate-related risks, thus potentially affecting the perceived insurability and risk assessments in insurance domains \citep{cerf2023participating}. The experts in this model addresses heavy-tailed distributions, drawing on the Extreme Value Theory (EVT) literature and its applications in insurance, as seen in the seminal works of \citet{Frechet1927}, \citet{Fisher1928}, \citet{Gnedenko1943}, \citet{deHaan1970}, \citet{balkema1974residual}. EVT has been applied to various aspects of climate change, including the analysis of extreme weather events \citep{coles2001introduction, palutikof1997economic} and the estimation of extreme financial losses \citep{mcneil2015quantitative}. The primary challenge, which remains unresolved, is that loss tails must be estimated with limited observations. Consequently, after an unforeseen tipping point, expert model uncertainty increases. This paper adds to the literature on modeling loss tails under uncertainty, building upon the works of \citet{danielsson2001using}, \citet{scarrott2012review}, and \citet{raftery2017less}, by exploring the combination of EVT and expert opinion.

This paper builds upon the literature on insurability and the challenges climate change poses to insurability. Climate change is testing the insurability of various business lines and geographic regions, as demonstrated by \citet{kunreuther1996mitigating}, \citet{jaffee1997catastrophe}, \citet{mills2007synergisms}, \citet{Charpentier:2008vn}, \citet{Kousky:2012wg}, and \citet{surminski2014role}. Both pandemic and climate change risks are difficult to mutualize in a cross-sectional manner due to their complex nature and widespread impacts. While pandemics have an episodic nature, insuring them intertemporally presents challenges due to their potential for widespread impact and external moral hazard \citet{richter2020covid}. Nevertheless capital markets (Cat bonds) and provisions can be efficiently used for pandemics with government promoting intertemporal behavior \citet{hartwig2020insurance}, a model for private pandemic insurance is suggested by \citet{grundl2021insurability}. Insurance companies cannot adopt the same strategies against climate change, as it is expected to be an irreversible trend with potential tipping points. This paper demonstrates that, although preventive actions against climate change are primarily the responsibility of regulators, a supervisor operating further down the line for insurance market stability can utilize this paper to direct his prudential activities to avoid disturbances caused by uncertainty. Our approach differs from the work of \citet{jaffee1997catastrophe} as we do not focus solely on catastrophes. \citet{jaffee1997catastrophe} consider that catastrophes are actuarially insurable as they are infrequent, local and uncorrelated.\footnote{\citet{jaffee1997catastrophe} considers that uninsurability is mainly due to the lack of (tax) incentives for insurance companies to accumulate a pool of liquid assets to meet catastrophe losses.} In our approach, we consider yearly cumulative losses each insurance company faces and how climate change fattens the subjectively estimated tail of this distribution up to uninsurability.

\subsection{Contribution Relative to Existing Research}

This paper makes three central contributions. First, it introduces the Pioneer Detection Method, a supervisory opinion-pooling framework specifically designed for environments in which climate change alters the underlying loss-generating process. Unlike traditional aggregation rules, the method identifies experts who detect structural shifts earlier than others and uses directional convergence to construct a supervisor-level estimate of the climate-sensitive tail parameter. This enables faster learning from fragmented, yearly aggregated loss data, particularly when extreme events are rare.

Second, the framework provides an operational supervisory tool for assessing insurability in markets where each insurer observes only a short private claims history. By weighting experts according to their ability to anticipate climate-induced changes, the Pioneer Detection Method allows supervisors to form a more robust and timely view of tail risks, even when reinsurance capacity contracts or granular data collection is limited.

Third, the method has broader applicability beyond insurance supervision. Because it formalizes how a system can identify and follow “pioneers’’ whose signals lead others during structural change, it can be adapted to contexts where multiple agents must jointly learn about a shifting environment. For instance, the approach could guide collective decision-making in autonomous systems, such as drone or vehicle swarms, where identifying agents that detect environmental changes earlier than their peers is crucial for stable and coordinated navigation.

Section \ref{sec:model} presents a model of the insurance market under climate change, where a risk-averse insurance buyer faces potential loss to an asset, and insurance companies operate in a competitive market with heterogeneous beliefs. The model also considers the effects of climate change on the tail parameter over time, the absence of reinsurance, and the role of an insurance supervisor who can influence the market by estimating and acting on the tail parameter. Section \ref{sec:tool} introduces the PDM, a novel approach to identify experts (pioneers) who deviate from the majority but toward whom other experts' opinions converge over time, especially in the context of rare events impacted by a changing climate. The method relies on three main steps, distance reduction, orientation change for convergence, and proportion of convergence attributed to a pioneer, and it also discusses alternative approaches. Section \ref{sec:experiment} details the testing and validation of the PDM against alternative opinion pooling methods in various scenarios, including insurance supervision under climate change and demonstrates that the PDM outperforms traditional measures in producing precise estimates earlier. Section \ref{sec:policyrecommendations} suggests that the decision to use the PDM in insurance markets depends on the market configuration and the evolution of the tail parameter. Through mathematical models and simulations, it demonstrates that in configurations with more than five insurance companies, the benefit of using the PDM depends on the cost of collecting information. However, in configurations with fewer than five insurance companies, it is always more beneficial to use the PDM to improve welfare, regardless of the cost function. Section \ref{sec:discussion} explores the complexities of policy recommendations for dealing with heterogeneous beliefs between insurance buyers and insurance companies in the context of climate change. Finally, Section \ref{sec:conclusion} summarizes the contributions to the fields of opinion pooling, climate change impacts, insurability, and insurance supervision, highlighting the introduction of the PDM as a practical tool to assess insurability and enhance market supervision amid climate change challenges; it emphasizes the method's speed and advantage in extreme events.

\section{A model of insurance market under climate change}\label{sec:model}

A risk-averse insurance buyer (IB) with an initial aggregated wealth $w$ and a von Neumann-Morgenstern utility function denoted by $U$, with $U'(.)>0$ and $U''(.)<0$, faces a risk of (yearly) aggregated loss of $x^t$ to his asset, where $x^t$ is a random variable following a Pareto distribution $\mathcal{P}\left(x^t, \alpha^t\right)$ with an unknown tail parameter $\alpha^t$ and a threshold normalized at unity.

We follow \citet{kleiber2003statistical} review on the usage of Pareto, a parsimonious model that is still highly effective for capturing the right-tail behavior of loss distributions. Insurance losses, especially in sectors like property, liability, or catastrophe insurance, often exhibit a heavy-tailed distribution where large, rare losses dominate the aggregate risk. The actuarial literature used the Generalized Beta 2 (GB2) family to model annual fire losses. In terms of likelihood, the full flexibility of the GB2 is not required and a one-parameter limiting case is sufficient that is related to a Pareto distribution after variable transformations. Insurance coverage above a deductible converge to the  Generalized Pareto Distribution (GPD) for given class of functions in \citet{balkema1974residual}. To calibrate the estimates we use the extreme value theory in \citet{embrechts2013modelling}.

An insurance contract for an insurance company (IC) $i$ is defined as a pair $\left( I\left(x^t\right), \Pi_i\left(I\left(x^t\right) \right) \right)$, with an indemnity schedule $I(x^t)$ and an insurance premium $\Pi_i\left(I(x^t) \right)$. Climate change, combined with macroeconomic factors such as inflation, impacts the tail parameter over time $t$ with potentially disturbing tipping points, but the realization of $\alpha^t$ is never observable. In this paper, both climate and economic effects on risk are referred to as ``climate change" for simplicity. Each IC is considered a Bayesian expert who estimates the tail parameter and calibrates its insurance contract based on public information and its private claim histories. We consider $m$ risk-neutral ICs that operate in a perfectly competitive insurance market without access to reinsurance. These ICs have homogeneous market shares for the asset class but potentially heterogeneous beliefs denoted by $\hat{\alpha}_i^t$.\footnote{We follow the notation of \citet{brunnermeier2014welfare} for beliefs heterogeneity: $\mathbb{E}_{\hat{\alpha}_i^t}\left(x\right) = \int x \mathcal{P}\left(x, \hat{\alpha}_i^t\right)dx$.} Each IC has its own private cost policy, denoted as $\gamma_i$. We make the usual assumption that $\gamma_i'(.)\geq 0$ to reflect the monitoring and auditing efforts made during the claims processing. There is no modeling of strategic behaviors of IC, and market shares are considered as fixed and exogenous. Hence, each IC offers the insurance premium Equation \ref{eq:inspremium}.

\begin{equation}
\Pi_i \left(\frac{1}{m} x^t \right)= \mathbb{E}_{\hat{\alpha}_i^t}\left[ I\left(\frac{1}{m} x^t\right) + \gamma_i \left( I\left(\frac{1}{m} x^t\right)  \right) \right]
\label{eq:inspremium}
\end{equation}

The IB is offered an aggregated premium $\Pi = \frac{1}{m} \sum_{i=1}^m \Pi_i$ for its asset, and we assume there exists a synthetic tail parameter belief $\bar{\alpha}^t$ such that $\Pi (x^t)= \mathbb{E}_{\bar{\alpha}^t}  \left[ I(x^t) + \gamma\left(I\left( x^t \right) \right) \right]$ where $\gamma$, as a sum of increasing functions, is increasing. The optimal insurance contracts are solution of the simplified program Equation \ref{eq:programeq}. There are two sources of beliefs heterogeneity in this model. The first source of belief heterogeneity comes from the heterogeneous claim histories that will drive the posterior IC belief $\hat{\alpha}_i^t$, but will not impact how program Equation \ref{eq:programeq} is solved compared with the literature. The second source of heterogeneity comes from the IB subjective probability $\hat{\alpha}_b^t$ which can differ from the IC belief $\bar{\alpha}^t$ and is the main point of attention on how program Equation \ref{eq:programeq} is solved. 

\begin{equation}\label{eq:programeq}
\begin{cases}
\max_{I(x^t),\Pi} W\left( I(x^t),\Pi \right):=\mathbb{E}_{\hat{\alpha}_b^t}\left[ U\left( w-x^t + I(x^t) -\Pi \right) \right]\\
\mbox{subject to } \Pi (x^t)= \mathbb{E}_{\bar{\alpha}^t} \left[ I(x^t) + \gamma\left[ I( x^t)  \right] \right] \\
\mbox{and }0 \leq I(x^t) \leq x^t
\end{cases}
\end{equation}

\citet{raviv1992design} extends \citet{arrow1963} and demonstrates that if there are no belief heterogeneity between the IB and the IC, an optimal contract is full insurance above a straight deductible.  \citet{gollier2013economics} shows that optimism on the side of the IB in the sense of a monotone likelihood ratio explains coinsurance above the optimal deductible. \citet{chi2019optimality} shows that full insurance over a straight deductible is always optimal if and only if the IC is more optimistic about the conditional loss given non-zero loss than the IB in the sense of monotone hazard rate order (MHR). We follow the MHR hypothesis for this paper, which allows for simple simulations. Without loss of generality, in the MHR case, We normalize this straight deductible to unity, which can later be multiplied by the order of magnitude of the asset class under consideration.  This simplifies the insurance contract to a one-dimension parameter problem where the indemnity follows a Pareto with a varying tail parameter $\hat{\alpha}_i^t$ and a unity threshold. The tail parameter is sufficient to characterize both the magnitude of the expected indemnity and the tail of the distribution: $\mathbb{E}^i\left[ I\left(x^t\right)\right] =\frac{\hat{\alpha}_i^t}{\hat{\alpha}_i^t-1}$.

If an asset class is extensively reinsured, information about realized losses is naturally centralized at the reinsurer. By aggregating a broad portfolio of claims, the reinsurer accelerates statistical learning about the tail of the loss distribution and thereby mitigates informational frictions. In such an environment, the supervisor has little need to resort to the PDM, unless reinsurers are unwilling to disclose their private assessments. The very presence of comprehensive reinsurance coverage may itself serve as a market signal, since reinsurers would not commit capital without having formed sufficiently precise beliefs about tail risks. For the purposes of our analysis, we therefore restrict attention to cases in which reinsurance is absent for the relevant asset class. This assumption is motivated by the fragility of reinsurance markets under climate change: when perceptions of tail risk shift abruptly at a tipping point, reinsurers may withdraw from exposed asset classes or sharply reduce their capacity. In these circumstances, the informational pooling provided by reinsurance disappears precisely when it would be most valuable. The supervisor must then rely on tools such as the PDM to generate credible estimates without developing in-house underwriting expertise or conducting costly on-site inspections.

An insurance supervisor, denoted as $S$ and mandated for ensuring financial stability, completes the model.\footnote{Real-life examples will be the International Association of Insurance Supervisors (IAIS)'s members: European Insurance and Occupational Pensions Authority (EIOPA), France's Autorit\'{e} de Contr\^{o}le Prudentiel et de R\'{e}solution (ACPR), Germany's Bundesanstalt f\"{u}r Finanzdienstleistungsaufsicht (BAFIN), UK's Prudential Regulation Authority (PRA), etc.} $S$ is a social planner whose program aims to maximize the welfare of the IB and IC, which is equivalent to maximizing the IB welfare by solving the program Equation \ref{eq:programeq}.\footnote{This is a ``benevolent supervisor assumption" which is equivalent to consider that $S$ dislike leaving a rent to the IC or considering $S$ maximizes IB and IC surplus when the IC is risk-neutral in a competitive insurance market.} Following \citet{brunnermeier2014welfare}, $S$ is aware of the presence of belief heterogeneity among IC but is unaware of the objective belief. Therefore, $S$ has to form his own tail parameter estimate, $\hat{\alpha}_S^t$. $S$ has two main options: 1) to exploit the estimates $\hat{\alpha}_i^t$ at a fixed cost $c_S$ and pool their expertise $\hat{\alpha}_{pool}^t$, with $c_S$ as a fixed cost for dedicating a team to this task based on available data) or 2) to audit IC (e.g. with on-site ad-hoc inspections) to estimate $\hat{\alpha}_{audit}^t$ after collecting detailed claim history at a cost $c_S + \lambda \# x$, with a term linear in the size of the information to collect $\# x$. As part of his supervision activities, $S$ already collects internal model reporting from IC and thus has access to their individual estimates $\hat{\alpha}_i^t$. We make two assumptions about how $S$ modifies the program Equation \ref{eq:programeq}. First, we assume that the functioning costs of $S$ are embedded in the costs of the IC. As ICs pay fees to the supervisor, these fees are passed on to the IB as part of the IC cost function. Secondly, we assume that $S$ announcement is fully trusted by an IC for their offered contracts.\footnote{For completeness, once calibrated, $S$ should integrate in the simulation how his announcement, $\hat{\alpha}_{S}^t$, can be integrated by an expert depending on his evaluation of his credibility and alignment with his objectives.} Equation Equation \ref{eq:intervention} formalizes the two impacts of possible $S$ interventions. Traditional opinion pooling methods have not been optimally designed for the EVT context. In an EVT context, extreme events do occur but with low probability, hence ICs will have heterogeneous learning speeds of the tail parameter, depending on whether their small samples contain extreme realizations. We design a Pioneer Detection Method for this context in Section \ref{sec:tool}.

\begin{equation}\label{eq:intervention}
\begin{split}
\mbox{Intervention after pooling: } \mathbb{E}_{\hat{\alpha}_{pool}^t}& \left[  U\left( w-x^t + I(x^t) -\Pi \left[ x^t \right]+ c_S  \right) \right]\\
\mbox{Intervention after audit: } \mathbb{E}_{\hat{\alpha}_{audit}^t}& \left[U \left( w-x^t + I(x^t) -\Pi \left[ x^t \right] + c_S + \lambda \# x   \right) \right]\\
\end{split}
\end{equation}

\section{A Pioneer Detection Method}\label{sec:tool}

The design of a new method for weighting expert estimates is informed by two primary assumptions of the model. First, estimations can never be compared to a realization of the tail parameter for weight calculation based on historical performance or forecast errors \citep{genest1986,stock2004combination}. As a result, the tool must rely on cross-sectional and temporal comparisons of expert model outcomes. Second, the distributions exhibit fat tails, indicating that extreme events occur infrequently but do happen. Experts learn from their claims; hence, experts already exposed to an extreme event at time $t$ will have faster learning rates compared with experts not yet exposed to tail events. Because insurers operate with proprietary datasets, heterogeneous learning speeds reflect sample differences rather than persistent misestimation. Bayesian updating ensures eventual convergence to the true tail parameter, although convergence is slow when losses are fat-tailed. In a climate-change environment, however, a subset of experts may internalize structural change earlier than others. The Pioneer Detection Method (PDM) aims to identify such pioneers as early as possible to accelerate learning.

\textit{Pioneers} are experts who deviate from the majority opinion, but other experts' opinions converge toward them over time, even though the experts do not cooperate.\footnote{\citet{liefferink2017} introduce a clear distinction between pioneers as being `ahead of the troops or the pack' and leaders which have `the explicit aim of leading others, and, if necessary, to push others in a follower position'.} This can also be thought of as implicit inter-temporal voting among experts to identify pioneers.

In section \ref{sec:alternativenovelapproaches}, we introduce alternative novel approaches in this context and will test them against the PDM. In Appendix \ref{sec:convergenceprop}, we study the convergence properties of the PDM.

The main benefit of the PDM is that the supervisor recognizes the convergence as soon as it begins and does not wait for the differences in estimates to narrow; a convergence in direction is enough to trigger a shift in weights. This PDM could also be envisaged for other contexts where non-cooperating experts are learning on rare events.

\subsection*{Initialization without Pioneer situations}

At time $t=0$, no past observation exists, hence the weights $w_i^0$ have to be initialized and can be taken as a linear combination, i.e. $w_i^0=\frac{1}{m}$.

There can be situations where no Pioneer is identified. A simple example is a case where there are only two experts and at time $t$ both are diverging from one another. In this case, the weights from the period $t-1$ can be replicated.

There can be some exotic situation where since the initialization period no Pioneer can be detected, hence the linear weights $w_i^0$ are still applied all along. This situation is not so much a weakness of the PDM, but rather a characterization of a situation where no implicit consensus emerges among experts. Taking a linear combination of diverging opinions might not be optimal and $S$ would need to investigate to understand why a persistent divergence can exist before building his own estimate.

\subsection*{Step 1: distance reduction condition}

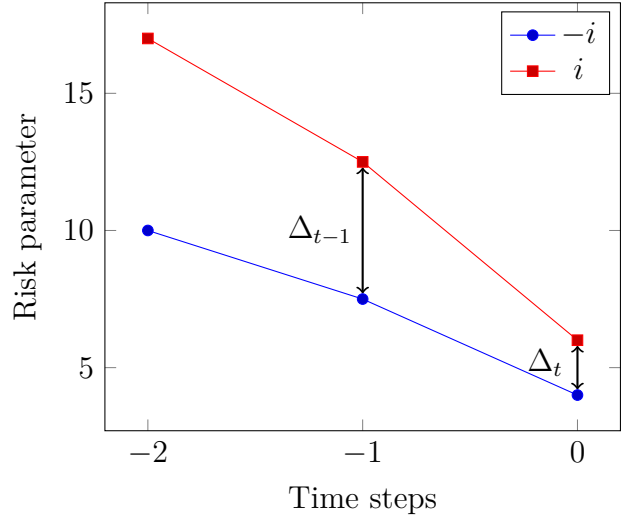
\begin{figure}[H]
\caption{Distance reduction dummy to identify Pioneers}\label{fig:step2}

\begin{tikzpicture} \begin{axis}[
    xlabel={Time steps},
    xtick={-2,-1,0},
    ylabel={Risk parameter},
    ]
    \addplot+[blue] coordinates {
        (-2,10) (-1,7.5) (0,4) 
    };
    \addplot+[red] coordinates {
        (-2,17) (-1,12.5) (0,6)
    };
    \legend{$-i$,$i$}
    \plot [<->, black, thick] coordinates {(-1,7.7)  (-1,12.3)};
    \plot [<->, black, thick] coordinates {(0,4.2)  (0,5.8)};
    
    \node[] at (axis cs: -1.2,10) {$\Delta_{t-1}$};
    \node[] at (axis cs: -0.15,5.2) {$\Delta_{t}$};
\end{axis}
\end{tikzpicture}
\begin{minipage}{.7\linewidth}
\footnotesize
\emph{Notes:} $i$ represents the expert of interest and $-i$ his competitors ($i$ excluded). $\delta_{\mbox{distance}}^t = \mathbbm{1}_{\Delta_{t} < \Delta_{t-1}} $
\end{minipage}
\end{figure}

First, we determine if the distance between an expert's estimate and the other experts' estimates has decreased between $t-1$ and $t$. This is illustrated Figure \ref{fig:step2} and represented by a dummy variable, $\delta_{\mbox{distance}}^t$, which serves as a necessary but not sufficient condition for an expert to be considered a pioneer. The other experts estimates can be defined as the average, but we also take the median and our results are unchanged.

\subsection*{Step 2: orientation change for convergence condition}
\begin{figure}[H]
\caption{Orientation change dummy to identify Pioneers}\label{fig:step3}

\begin{tikzpicture}[scale=0.7] \begin{axis}[
    xlabel={Time steps},
    xtick={-2,-1,0},
    ylabel={Risk parameter},
    ]
    \addplot+[blue] coordinates {
        (-2,17) (-1,14) (0,9)
    };
    \addplot+[red] coordinates {
    (-2,10) (-1,7.5) (0,5) 
    };
    \legend{$-i$,$i$}
    \addplot+[red, dashed] coordinates {
        (-1,7.5) (0,7.5) 
    };
        \addplot+[blue, dashed] coordinates {
         (-1,14) (0,14)        
    };
    \coordinate (A) at  (0,14) ;
    \coordinate (B) at (-1,14);
    \coordinate (C) at (0,9);
    \pic [draw, ->, "$\theta_{-i}^t$", angle radius=1.7cm, angle eccentricity=1.5] 
    {angle = C--B--A};
    \coordinate (D) at  (0,7.5) ;
    \coordinate (E) at (-1,7.5);
    \coordinate (F) at (0,5) ;
    \pic [draw, ->, "$\theta_{i}^t$", angle radius=1.7cm, angle eccentricity=1.5] 
    {angle = F--E--D};
\end{axis}
\end{tikzpicture}
\qquad
\begin{tikzpicture}[scale=0.7] \begin{axis}[
    xlabel={Time steps},
    xtick={-2,-1,0},
    ylabel={Risk parameter},
    ]
    \addplot+[blue] coordinates {
        (-2,17) (-1,12.5) (0,6)
    };
    \addplot+[red] coordinates {
        (-2,10) (-1,2) (0,4) 
    };
	\legend{$-i$,$i$}
    \addplot+[red, dashed] coordinates {
        (-1,2) (0,2) 
    };
        \addplot+[blue, dashed] coordinates {
        (-1,12.5)  (0,12.5) 
    };
    \coordinate (A) at  (0,4) ;
    \coordinate (B) at (-1,2);
    \coordinate (C) at (0,2);
    \pic [draw, ->, "$- \theta_{i}^t$", angle radius=1.5cm, angle eccentricity=1.5] 
    {angle = C--B--A};
    \coordinate (D) at  (0,12.5) ;
    \coordinate (E) at (-1,12.5);
    \coordinate (F) at (0,6) ;
    \pic [draw, ->, "$\theta_{-i}^t$", angle radius=1.5cm, angle eccentricity=1.5] 
    {angle = F--E--D};
\end{axis}
\end{tikzpicture}

\begin{minipage}{.7\linewidth}
\footnotesize
\emph{Notes:} $i$ represents the expert of interest and $-i$ his competitors ($i$ excluded). The measure is $\delta_{\mbox{orientation}}^t = \mathbbm{1}_{\theta_{-i}^t > \theta_{i}^t}$. In both panels, $i$ can be considered a pioneer. The right panel is interesting because $i$ is not consistent over time, but is still considered a pioneer.
\end{minipage}
\end{figure}
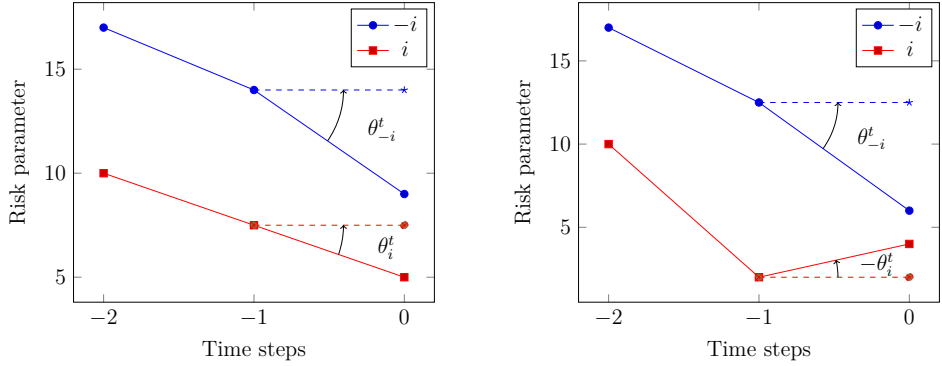

Second, we determine whether the orientation of the segments between an expert's estimate and the average of the other experts' estimates has decreased between $t-1$ and $t$ and whether this decrease can be attributed to the expert being considered a pioneer by his peers. An expert is considered a pioneer by his peers when the significant convergence of trends is due to the other experts agreeing with him. \emph{Agreeing} here means that the other experts estimate changes direction over time towards the expert's estimate. An expert is considered a follower when his estimate is the one converging toward the average of the other experts' judgments. This is illustrated Figure \ref{fig:step3} and represented by a dummy variable, $\delta_{\mbox{orientation}}^t$.

\subsection*{Step 3: proportion of the convergence attributed to a Pioneer}

Finally, if both step 1 and 2 dummies are True, we calculate the proportion of the orientation decrease attributable to the average of the other experts' estimates, $-i$, with respect to the considered expert's estimate, $i$. The pioneer score for each expert, indexed by $i$, is then derived from a combination of the four steps, as shown in Equation Equation \ref{eq:foursteps} and is used as a weighing for his opinion in the final estimate of $S$: $\hat{\alpha}_S^t=\sum_i w_i^t \hat{\alpha}_i^t$.
\begin{equation}\label{eq:foursteps}
w_i^t = \delta_{\mbox{distance}}^t \times \delta_{\mbox{orientation}}^t \times \frac{\vert \theta^t_{-i}\vert }{\vert \theta^t_{-i}\vert +\vert \theta^t_{i}\vert }
\end{equation}
where $-i$ represents the other experts, $i$ excluded.

\subsection*{A note on data quality}

The treatment of raw time series data may require preliminary data-quality processing, depending on the properties of the empirical signals under study. Human experts do not consistently behave in a Bayesian manner when reporting model calibrations, and their reported parameters can be affected by idiosyncratic noise and occasional errors. In applied settings, it may therefore be advisable to implement data-quality procedures aimed at improving the signal-to-noise ratio, thereby limiting the risk that observed trends reflect noise rather than genuine dynamics. In the context of this paper, however, such procedures are unnecessary, as the signals we generate are, by construction, derived from Bayesian experts.

\section{Testing and validating the Pioneer Detection Method}\label{sec:experiment}

We test and validate the PDM against alternative opinion pooling methods, using the Root Mean Square Error (RMSE) as a criterion to evaluate the precision of the estimate. We will use time series settings and evaluate the capacity of the PDM to produce a precise estimate earlier than traditional opinion pooling methods. We demonstrate the importance of this early capacity in the context of insurance supervision under climate change. In a severe climate stress scenario, the estimated tail of the yearly aggregate losses of an asset class can jump and reinsurance companies can decide to leave leave an asset class market. When most reinsurance companies are foreign and out of supervision and regulation scope of $S$, this leave little room for $S$ to influence the reinsurance sector on this asset class. Insurance experts are left alone to decide at which cost they can offer insurance coverage and experts can have heterogeneous beliefs. We model heterogeneous beliefs with Bayesian experts estimating the tail parameter based on their private samples.\footnote{It is common practice that insurance experts leverage also on public information and consulting companies or model provider can reduce the heterogeneity. While reliance on common model might raise other supervision concerns, we limit our modeling to estimates based on private information.} We evaluate the capacity of the PDM to help $S$ leverage on a limited count of insurance expert estimates and a limited time series history (two to five years) to decide on optimal supervision and regulation actions.

We start with the seminal data set on combination forecasts as a preliminary check. We find that the PDM produces coherent results and outperforms traditional measures. This first part is not for drawing a general conclusion; it serves as a sanity check to ensure that the PDM, if applied in a non-EVT context, would still perform. For the main validation, we model an insurance market with Bayesian experts who have potential heterogeneous beliefs, learning a fat tail parameter from private samples taken from the same loss EVT distribution. Across simulations, to validate the relevance of the PDM in different market settings, we vary the tail parameter and the number of experts to confirm that the PDM performs best, as early as possible, in any configuration. We cannot determine ex ante which experts will be identified as pioneers and to which extent, which will determine their weight over time, hence we will ultimately have to use Monte Carlo simulations with a known fat tail distribution for each simulation. We use a Pareto distribution as we argued in Section \ref{sec:model} that it covers generic situation of aggregate yearly losses with fat tails.

\subsection{Comparing the Pioneer Detection Method with Existing Combination Methods}

We compare the PDM with existing combination forecasting methods. Following the work of \citet{stock2004combination}, which relies on the dataset from \citet{bessler1981forecasting}, all methods are evaluated using their RMSE relative to the autoregressive model of order one calibrated at each period on all past periods.\footnote{An autoregressive model of order one, AR(1), for a stationary time series $\alpha^t$ is the simplest form $\alpha^t= \phi \alpha^{t-1} + \nu^t$ with $\phi$ a constant that can be calibrated with ordinary least squares method and $\nu^t$ an error term. If a more complex model cannot beat the AR(1), then this complex model may be discarded in favor of the parsimonious AR(1) \citep{box2015time}.} As with the original paper, we find that combination forecasting improve upon autoregressive forecasts (Table \ref{tab:combinationcompare}). The PDM performs best, although with only three forecasters, there are some situations where no expert can be identified as a pioneer (e.g., when all three experts are diverging from one another). This is a limitation in the application of this method for combining forecasts. Note that the three novel methods introduced—Granger Causality, Lagged Correlation, and Pioneers—do not rely on past performance and do not include the actual true time series to be estimated. This is a desired feature when climate is constantly changing and the true parameter of the EVT losses can never be exactly observed.

\subsection{Validation of the Pioneer Detection Method with Fragmented Insurance Market Shares}

We now want to evaluate the capacity of the PDM to ``learn as fast as possible" a loss distribution after a unique and unexpected change. For this test, we will use configurations with limited count of insurance experts $m \in [2,20]$ and tail parameter after the change around the problematic range $\alpha \in [0.5,3]$. We chose this range as for a Pareto loss distribution, as soon as the tail parameter is below $2$ the variance is unbounded and as soon as the tail parameter is below $1$ the mean is unbounded.

We evaluate the PDM on a fragmented insurance market through the use of Monte Carlo simulations. In a hypothetical steady state, IB and IC would agree on an estimation of $\alpha$, $\alpha^-$, although there can be heterogeneous beliefs among IC. This is equivalent to modeling that each IC would have his own estimates $ \hat{\alpha}_i^-$, but the representative offered insurance contract in the steady states follows $\alpha^-$. IB are not explicitly modeled in the simulations but are assumed to solve the program Equation \ref{eq:programeq} for the optimal insurance contract. We start the simulations with Bayesian experts who have uninformative priors and allow for two time periods, $t\in[0,1]$, as burn-in to account for beliefs' heterogeneity. At $t=2$ a climate change enters a unique and unexpected tipping point. The unknown tail parameter jump to a new value following the climate tipping point and we simulate that after this tipping point the climate has stabilized to an unobservable state, hence $\alpha$ remains stable. In this controlled environment, we can calculate the RMSE for each expert estimation against the true $\alpha$, after time $t=2$. We benchmark the performance of the PDM against existing opinion pooling techniques as well as other novel approaches introduced in this paper.

We model Bayesian insurance experts who learn losses generated from a Pareto Type I distribution with a tail parameter close to unity, indicating a fat-tail environment. Each expert's observations are independent from one another both in cross-section\footnote{Although each data generating process (DGP) is governed for each expert by the same $\alpha$, the observation of a given expert at time $t$ does not impact the value of the observation of another expert at the same time $t$. In other words, expert $i$ can observe an extreme event at time $t$, but this does not change the probability that expert $j$ observes an extreme event at the same time $t$.} and over time\footnote{The tail parameter remains stable during the learning phase, eliminating the need to model 'aiming for a moving target.' While in reality, climate change may not have a single one-time impact, here, we aim to evaluate the capacity of the PDM to rapidly adapt following a tipping point either in the true risk distribution or in the experts' perceptions.}. In such an environment, experts will observe extreme values over time, but might have to wait for some time before observing an extreme event and being able to refine their calibration of the tail parameter. The PDM starts with at least one set of past data points, and it weights pioneers as soon as a change of direction to their estimates occurs. The supervisor needs at least one past period of estimations to form his own subjective opinion with the PDM. Once the estimation can be formed, the PDM outperforms the other opinion pooling methods (Table \ref{tab:validatemeasure}). The performance of the new method is significant in early stages, especially at the first period where an estimate can be formed ($t=2$). The performance is also found to be more stable over time, using mean, median, and the standard deviation over the first ten periods (Table \ref{tab:stablemeasure}). An interesting challenger is to take the minimum tail parameter at each step. With heterogeneous beliefs, this can lead to taking the most conservative premium, but has the benefit of being a stable approach (Table \ref{tab:stablemeasure}, standard deviation). If subjective (non-Bayesian) experts judgments are added to the tail parameter estimate in a crisis, then taking the minimum estimate would exacerbate the \emph{panic }or \emph{run} on the insurance coverage provision.

We test the capacity of the new method to identify linear relationships between time series (Section \ref{sec:linearts}). As expected from the literature, in a Gaussian context, the linear opinion pooling performs best, and the new PDM does not improve performance, but its performance converges in the long term with the linear method (Table \ref{tab:validatetransform}). This performance is robust to scaling and non-linear transformations such as logarithmic (Table \ref{tab:validatetransform}), the method is not scale but ordinally invariant. 

Table \ref{tab:robustnesschecks} reports robustness checks for the new supervision tool where the tail parameter $\alpha$ and the number of Bayesian experts are varied. The PDM minimizes the RMSE for all configurations.

\section{Policy recommendations}\label{sec:policyrecommendations}

The recommendation to use the PDM for insurance supervision under climate change depends on the insurance market configuration and the evolution of the loss distributions. We use a logarithmic utility function as in \citet{mossin1968aspects} and apply the Delta method to estimate the gain or loss in utility at the $95\%$ lower confidence interval bound.\footnote{The logic is the same as the mean-variance principle for the premium applied by the IC, following \citet{young2014premium}, $\frac{\hat{\alpha}^t}{\hat{\alpha}^t-1} + \beta \frac{\left( \hat{\alpha}^t\right)^2}{t}\frac{1}{\left(\hat{\alpha}^t-1 \right)^4}, \quad \hat{\alpha}^t>1$, for which the variance coefficient would need to be determined in an experiment. \citet{cao2022stackelberg} demonstrate that in a Stackelberg competition, the ambiguity aversion determines this coefficient. For consistency, we use the confidence interval, where the representative contract offered by the IC assures they are profit-making $95\%$ of the time.} The loss of utility is proportional to the variance of the estimate of the tail parameter $\alpha$.\footnote{It is expressed by applying a Taylor expansion to $Var\left( \log\left[ c-  a \frac{\alpha}{\alpha-1} \right] \right) \simeq \left( \frac{a}{(\mu_\alpha-1)\left[ c(\mu_\alpha-1)-a \mu_\alpha \right]}\right)^2 Var(\alpha)$ around the mean $\mu_\alpha$ of the tail parameter with $c$ and $a$ being constants.} %I normalize the variance with the Bayesian estimate variance in the case of a market with a unique expert, then I can express the variance for the fully informed $S$ as $\frac{1}{m t}$ using expression Equation \ref{eq:posteriorexpert}.

\begin{figure}
\caption{Cost versus utility benefit of $S$ actions}\label{fig:CostvsCI}
	\includegraphics[width=0.5\textwidth]{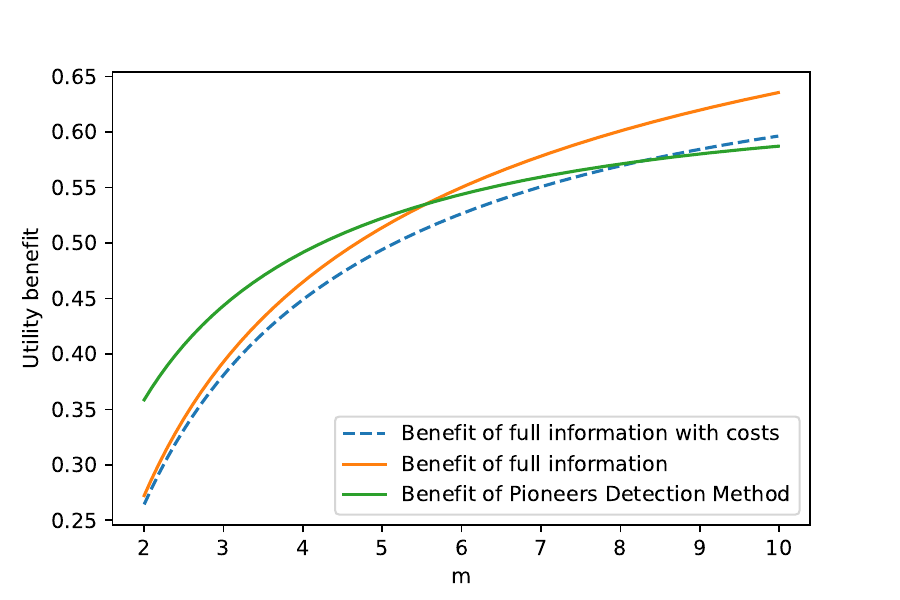}  
	\begin{minipage}{1\textwidth}
	{\footnotesize{Source: author's computation.}}
	\end{minipage}
\end{figure}

As there is no closed form for the variance of the estimate outcome with the PDM, we apply Monte Carlo simulations with a tail parameter $\alpha=1.5$. We vary the number of ICs, $m$, and find a strong linear relationship between $m$ and the ratio of the estimate standard deviations, $ \scriptscriptstyle \frac{\sigma \hat{\alpha}_{\text{tool}}}{\sigma \hat{\alpha}_{\text{full}}} = a + b m$. Figure \ref{fig:CostvsCI} illustrates the utility improvement versus the cost $S$ has to spend. As the number of experts increases and as long as they all face independent and identically distributed DGP, the benefit of the PDM to identify with limited observations which expert calibrated early the tail parameter vanishes as a sample with more observation starts to be more precise with progressively enough observations in the tail. In Figure \ref{fig:CostvsCI}, we display the normalized benefit of full information without the linear cost and find that $S$ would never find it beneficial to spend the effort to collect the full information set from IC using on-site inspection when there are fewer than five ICs. When $S$ incurs a linear cost to collect granular information, this threshold shifts to the right, we illustrate this with a dashed line, although the cost function would need to be calibrated on real data. 

%With my model and setup, in a configuration with more than five ICs, whether it is welfare-improving for $S$ to collect the full set of information to reduce the uncertainty in the premium estimation depends on the cost to collect granular information. In a configuration with less than five ICs, it is always more beneficial for $S$ to use the PDM, regardless of the cost function, to improve the welfare.

\section{Discussion}\label{sec:discussion}

\subsection{On the existence and persistence of pioneers in a competitive insurance market}

In Section \ref{sec:model}, a formalized framework was presented, depicting a competitive insurance market with Bayesian experts. This section aims to open the road for future work testing the applicability and emergence of pioneering strategies within such a market in real-world scenarios, where insurance experts might not apply Bayes' rule and where competition might interfere in their judgment and dynamically modify market shares.

Let's assume first that insurance expert are all Bayesian, the PDM is only overperforming traditional opinion pooling methods in an extreme value (EV) context. In a Gaussian context, even with a limited sample size (fewer than ten observations), a Bayesian expert's first and second moment estimations will be unbiased. Therefore, in a Gaussian context, the PDM wouldn't outperform a linear combination that excludes outliers. One can infer this intuitively recognizing that each observation taken out of a Gaussian DGP is equally likely to lie to the right or the left of the true first moment. This resembles the law of large numbers, where the average of means is itself Gaussian-distributed. However, when dealing with EV distributions, the law of large numbers isn't applicable when the tail index implies an unbounded mean or variance and furthermore a sample of small size might not be representative the distribution's tail, as tail events are rare and might not be reflected in a small sample. Intuitivel, in an EV context, a supervisor relying solely on a linear combination of tail indices might be misguided and I demonstrated in this paper that with Bayesian experts and fixed market shares the PDM is overperforming. 

Yet, if systematic bias emerges, for example after a tipping point, the situation changes fundamentally. As \citet{min1993bayesian} emphasize, the presence of bias can render forecast combinations less accurate than individual forecasts, thereby eliminating the efficiency gains from averaging. This holds in particular when the bias is systemic, that is, when the distortions of one expert do not offset those of others. In such cases, where all experts consistently overestimate or underestimate tail risk, neither the PDM nor any traditional opinion pooling method can correct for this distortion. Methods such as those proposed by \citet{giacomini2009detecting} provide formal statistical tests to detect forecast breakdowns and thus systematic bias. However, since the tail parameter is not directly observable to the supervisor, such tests cannot be applied in the context of insurance under climate change. In practice, only on-site inspection of insurers' internal data and models would enable the supervisor to identify and address systemic bias in tail risk assessments.

A significant body of experimental research \citep{griffin1992weighing, antoniou2015subjective} indicates a prevalent deviation from Bayesian reasoning among individuals, which may apply to insurance professionals. This deviation raises questions about the robustness and effectiveness of the PDM in these settings. Consequently, it suggest further empirical investigation with a controlled experiment\footnote{We thank Nicolas Jaquemet, Lily Savey and the anonymous Paris School of Economics Institutional Review Board for their advice on this future experiment project.}, wherein actuaries are engaged in risk evaluation tasks under conditions of limited information, within both Gaussian and EV distributions. We would test if human experts adhere to Bayesian principles or whether they incline towards excessive reactions, setting \textit{exuberant} premiums. If such overreactions are identified, it could challenge the PDM's current framework, potentially necessitating modifications such as the introduction of inertia in pioneer identification to prevent overshooting. We conducted a preliminary uncontrolled assessment leveraging on an end-of-year examination administered to master's level students at University Paris 1 in 2021. The informal findings open the possibility to study further the behavior of experts in a rigorous experimental design, not just to validate the PDM but also to test the underlying assumption of Bayesian agents and competition.

The critique that the PDM may primarily be effective in markets with a limited number of insurers, specifically ranging from 5 to 8, necessitates further discussion. While it's acknowledged that the insurance market, notably in countries like France, is diverse and expansive, there are indeed niche markets where only a select few insurers offer coverage for specific asset classes. A pertinent example is the French local authorities' insurance market in 2023, where challenges in securing coverage were exacerbated by a limited number of insurers, highlighting a potential application scenario for the PDM. The observation raises valid concerns regarding the method's generalizability and practical relevance. The effectiveness of the PDM, in essence, hinges on the cost of data collection and structure of the market it is applied to, rather than the sheer number of participating insurers. Future extensions of this work could include a model for conducting a cost-benefit analysis of information collection that takes into account the complexity of data, the number of events to aggregate, and the availability of technological tools for data analysis.

Turning our attention to the French insurance market overseen by the ACPR, AXA SA stands out as the predominant player. This insurer has even carved out a dedicated entity, AXA Climate, which has ventured into consultancy. Evidently, size heterogeneity and market share play pivotal roles, enabling certain players to establish specialized teams focused on discerning climate change implications for the insurance sector. While we refrain from commenting on the commitment of any specific French firms in climate change modeling, size disparities naturally pave the way for the birth of pioneers. Since market shares are fluid, it's conceivable that a pioneering insurer might be better poised to make informed decisions regarding which asset classes to insure or exclude ahead of competitors. Such dynamics could skew the insurance market. While these facets aren't encapsulated in our current model, they could be fertile ground for future experiments, especially in exploring how markets might evolve over time under the influence of EVT risks. In this paper's design of the PDM, the weights are a ratio of angles of changes between the previous period and the current observation's period, in other words there is no inertia in our main specification of the PDM. This can be updated based on equation Equation \ref{eq:foursteps} where the weight for each expert becomes $\sum_{k=0}^l \theta^k w_i^{t-k}$, now the optimal lag order ($l$) selection becomes an additional research question that cannot rely on the traditional time series approach \citet{helmut2005} as the supervisor has no access to the realization of the tail parameter for calibration. An other extension would be to consider that now the climate change has a frequent impact on the tail parameter and the experts are trying to internalize this an model: $\hat{\alpha}_i^t = f\left( \alpha^{t-1}\right) + \epsilon_i^t$.

The speed of learning is naturally related to market share. Under Bayesian updating, a predominant insurer emerges as a natural pioneer, since its larger portfolio accelerates inference about tail risks. By contrast, if the predominant insurer follows a non-Bayesian, more traditional approach and ceases to update beliefs effectively, the PDM can discount this ``inertial player'': market share carries no explicit weight in the mechanism, either contemporaneously or historically. The PDM is an inter-temporal silent voting scheme that only the supervisor with its data can establish, and remains untainted by insurers' relative size. However, the mechanism fails if the majority of insurers do not learn. In such circumstances, systemic risk escalates: true risk rises sharply while premiums remain stable. The result is that most insurers incur losses and face exit or are forced into belated, accelerated learning. Once learning resumes, the PDM again becomes a reliable tool for the supervisor.

A further limitation arises when insurers' portfolios differ substantially in their geographical exposure. In such cases, the underlying tail parameters of the loss distribution are no longer homogeneous across firms. If the supervisor were to combine estimates that implicitly refer to different tail parameters, this would create a misspecification problem. The PDM, by construction, is designed for a univariate setting where the risk class is common across insurers. In practice, however, climate-related risks often exhibit strong regional heterogeneity, for example, coastal real estate in Dunkerque, where sea-level rise poses particular modeling challenges. In such contexts, the appropriate application of the PDM is to restrict pooling to the subset of insurers exposed to the same peril within the same region. Applying the method indiscriminately across heterogeneous asset classes or geographical exposures would invalidate not only the PDM but also any conventional pooling approach, as the estimates would not be commensurable.

\subsection{On the role of leader of a supervisor}

This paper has delineated policy recommendations, highlighting configurations in which a supervisor can employ the PDM for risk estimation to inform decisions pertaining to intervention. Equation \ref{eq:intervention}, alongside computations considering welfare change (either gains or losses), encompasses supervisory interventions as a leader that either directly or indirectly disclose the supervisor's own estimations of the tail parameter, thereby acting as an indicative benchmark of the most astute available estimate at the time of disclosure. This evokes a pivotal inquiry into the incentives for private insurance companies to amplify their expertise when confronting the prospect that their discernment will subsequently be utilized by a supervisor, synthesized through a PDM, and disseminated amongst all insurance entities.

This aforementioned trade-off is not rigorously operationalized in the present work and somewhat parallels earlier discourses concerning the potential veracity of the \textit{Efficient Market Hypothesis} (EMH). The hypothesis postulates that financial markets are efficient in reflecting all available information in asset prices. This brings forth the Grossman-Stiglitz Paradox, which presents a conundrum: if every piece of information is already embodied in prices, the motivation for any investor to incur costs to acquire information becomes obfuscated \citep{grossman1980impossibility}. Analogously, in the context of insurance, if supervisors utilize and disclose expert risk estimations via interventions, it could ostensibly dampen insurance companies' incentive to enhance their own expertise. Essentially, why would insurance companies bear the costs of accruing and refining expertise if supervisory interventions eventually render such proprietary insights into public knowledge?

\subsection{Absence of reinsurance in the model}

A central hypothesis in this model is the absence of any form of reinsurance, neither from outside reinsurers nor between ICs. This simplifying assumption can be understood as a situation where after a tipping point for an asset class in a country supervised by $S$, foreign reinsurance companies (not under direct and unique $S$ supervision) decide to exit the market or provide deterrent reinsurance premium during the January renewal. This is a strong modeling assumption with limited historical precedent such as the liability insurance crisis of the mid-1980s, due to underestimation of long-tail liabilities, legal and macroeconomic (interest rates) changes \citep{berger1992reinsurance}. As an anecdotal recent evidence post COVID-19, on 8th December 2020, the reinsurer Munich Re announced it would stop covering big event cancellation.

The Californian wildfires of 2025 further illustrate this dynamic. According to Citi analysts, reinsurers absorbed less than 3\% of insured losses from these fires, as they had significantly reduced their exposure to wildfire risks in recent years \citep{ft2025wildfires}. This retreat left frontline insurers and homeowners heavily exposed, with many turning to the California state-backed Fair Plan or remaining uninsured. Similarly, in France, where climate change has intensified drought risks, if WSCS were excluded from state guarantees \citep{cc2022}, reinsurers might exit certain areas, forcing ICs to determine premiums independently.

As a simplification, neither dynamic entry nor exit of reinsurers or ICs is allowed before or after the tipping point. This is an interesting avenue for extension of the model where $S$ can influence entry or exit of (re)insurers depending on his own evaluation of the risk-benefit assessment.

According to \citet{plantin2006}, insurers rarely trade risks with each other. In his model, reinsurers emerges as informed capital providers that closely monitors IC to mitigate moral hazard problem because they have granular (and soft) claim information, and more risk-management skills than outside financiers. Once reinsurers exit a given market, $S$ can decide to take over this monitoring role and start collecting granular claim information or use opinion pooling method to synthesize IC expertise.

\section{Conclusion}\label{sec:conclusion}

This paper contributes to the literature on insurability and insurance supervision in the context of climate change. By introducing a Pioneer Detection Method that pools expertise, the paper offers a practical approach to assess the insurability of asset classes after a tipping point. It tests the policy recommendation that insurance supervisors should focus on on-site inspections to collect granular information or should rely on the pooling of expertise. In the event of a sudden shock or panic in the insurance market for a given asset class, a supervisor must determine as quickly as possible whether the asset class is insurable to fulfill his stability mandate. The new tool relies on readily available model parameters to identify experts who are quick at understanding the implications of climate change on aggregate tail losses. We demonstrate that, compared to traditional opinion pooling methods, the Pioneer Detection Method is the fastest and has an advantage when there are limited observations available. The choice to collect granular information via on-site inspection will depend on the supervisor's capacity and his estimate of the cost-benefit. We demonstrate that, as long as there is a limited number of independent insurance companies covering an asset class, relying on the Pioneer Detection Method is always preferred.

In terms of policy implications, after building their own estimates, supervisors can reduce uncertainty and help avoid crises and equilibrium shifts in insurance supply by making forward-looking announcements and influencing expert opinions. A supervisor can also use the Pioneer Detection Method to monitor the insurability of asset classes and advise a regulator to design appropriate public or private insurance schemes. Additionally, it suggests that transparency and collaboration within the insurance sector can contribute to a more resilient market in the face of climate change.

Further research on the behavior of insurance experts should follow, especially to test if they exhibit an optimism or pessimism bias due to the lack of extreme observations in their private datasets, or because of modeling choices. We recommend setting up experiments to determine how actuaries would behave in this context, particularly to test whether their approaches are Bayesian. The paper also highlights the need for further research in the area of insurance supervision and climate change, exploring topics such as the endogeneity of market shares, demand sensitivity to premiums, and ruin gambles in the context of a changing climate. 

%Future studies could investigate the effectiveness of the proposed tool in different regulatory environments and under varying degrees of climate uncertainty, refining the tool and providing valuable insights for insurance supervisors and regulators worldwide.

%In conclusion, this paper presents a novel supervision tool for pooling expertise under radical uncertainty, with significant relevance in the context of climate change. The proposed tool can assist insurance supervisors and regulators in better assessing the insurability of asset classes, maintaining financial stability, and designing effective public and private insurance schemes. 
%By promoting transparency and collaboration within the insurance sector, the new tool can contribute to a more resilient insurance market better equipped to face the challenges of climate change. This paper underscores the importance of continued research in this area, as the impacts of climate change on the insurance sector continue to evolve and grow in significance.

 	%references
 	\newpage
	 \bibliographystyle{aea}
	 \bibliography{MLreass}
	
	% figures
	%\input{figures}
	
	% tables

% table
\begin{table}[!htbp] \centering
\caption{New supervision tool compared with combination methods}\label{tab:combinationcompare}
\begin{tabular}{lrllll}
 & RMSE & Constant & Econometric & ARIMA & Expert \\
Expert & 1.35 &  &  &  &  \\
Econometric & 1.22 &  &  &  &  \\
Method C out-sample & 1.00 & -22.21 & 0.93 & 0.22 & 0.31 \\
AR & 1.00 &  &  &  &  \\
Method A out-sample & 0.79 & 0 & -1.11 & 4.63 & -2.62 \\
Method B out-sample & 0.69 & 0 & -1.01 & 4.63 & 0.28 \\
ARIMA & 0.64 &  &  &  &  \\
Median & 0.64 &  &  &  &  \\
Minimum MSE adaptive & 0.61 &  & 0.32 & 0.41 & 0.27 \\
Discounted MSFE & 0.57 &  & 0.19 & 0.46 & 0.35 \\
Minimum MSE adaptive, expert excluded & 0.57 &  & 0.41 & 0.59 &  \\
Correlation & 0.56 &  & 0.30 & 0.35 & 0.36 \\
Two forecast Simple average & 0.55 &  & 0.50 & 0.50 &  \\
GC & 0.55 &  & 0.50 & 0.50 & 0.00 \\
Minimum Variance, expert excluded & 0.55 &  & 0.48 & 0.52 &  \\
Three forecast Simple average & 0.52 &  & 0.33 & 0.33 & 0.33 \\
Minimum Variance & 0.48 &  & 0.53 & 0.25 & 0.22 \\
Method B & 0.47 & 0 & 0.54 & 0.25 & 0.22 \\
Method A & 0.47 & 0 & 0.53 & 0.25 & 0.22 \\
Method C & 0.44 & 15.19 & 0.10 & 0.40 & 0.18 \\
Pioneers & 0.42 &  & 0.70 & 0.17 & 0.12 \\
\end{tabular}

\Tabnote{The data set is taken from \citet{bessler1981forecasting}. The reported weights are the average when they are varying over time for the method. Simple averages, Minimum MSE adaptive and Minimum Variance are implemented as in  \citet{bessler1981forecasting}. Methods A, B and C are from \citet{granger1984improved}, weights were computed on past performances for the out-sample. Discounted MSFE is the first method in \citet{Bates:1969vx}. Principal Component forecast combination was not added as this is mainly to tackle situations with a large number of forecast. I add three novel methods, Granger Causality (GC), (lagged) Correlation and the Pioneer Detection Method.}
\end{table}

% table
\begin{table}[!htbp] \centering
\caption{New supervision tool validation - full Bayesian}\label{tab:validatemeasure}
\begin{tabular}{lrrrrrrr}
 & Pioneers & Linear & Median & Minimum & Pioneers distance-weighted & Correlation & GC \\
2 & 1.00 & 6.57 & 2.46 & 1.13 & 1.83 & 792.00 & 8.18 \\
3 & 1.00 & 5.94 & 2.00 & 1.27 & 1.68 & 315.40 & 6.93 \\
4 & 1.00 & 2.84 & 1.68 & 1.52 & 1.22 & 87.38 & 3.75 \\
5 & 1.00 & 2.10 & 1.52 & 1.69 & 1.05 & 391.41 & 2.83 \\
6 & 1.00 & 1.72 & 1.39 & 1.76 & 0.95 & 62.55 & 2.42 \\
7 & 1.00 & 1.48 & 1.29 & 1.80 & 0.89 & 46.78 & 2.17 \\
8 & 1.00 & 1.31 & 1.20 & 1.81 & 0.84 & 51.38 & 1.97 \\
9 & 1.00 & 1.18 & 1.13 & 1.79 & 0.81 & 53.14 & 1.82 \\
\end{tabular}

\Tabnote{The tail parameter is taken from a Pareto type one distribution with $\alpha=1.5$. Five non-cooperative bayesian experts are modeled with independent observations from the loss distribution, heterogeneous beliefs are allowed. $10^5$ Monte Carlo simulations are run. The new Pioneer Detection Method outperform established opinion pooling methods (linear, median and minimum) as well as alternative methods introduced in this paper: pioneer with weight based on distance rather than angle, lagged correlation and Granger Causality.}
\end{table}

% table
\begin{table}[!htbp] \centering
\caption{New supervision tool validation - full Bayesian}\label{tab:stablemeasure}
\begin{tabular}{llllllll}
 & Pioneers & Linear & Median & Minimum & Pioneers distance-weighted & Correlation & GC \\
mean & 1.00 & 3.62 & 1.74 & 1.51 & 1.29 & 172.00 & 4.59 \\
median & 1.00 & 1.92 & 1.46 & 1.72 & 1.00 & 85.12 & 2.64 \\
std & 1.00 & 10.34 & 3.16 & 0.71 & 2.50 & 125.00 & 12.34 \\
\end{tabular}

\Tabnote{The tail parameter $\alpha$ is fixed at $1.5$ and a Monte Carlo simulation is run with $10^5$ runs. Five non-cooperative Bayesian experts are modeled with independent observations from the loss distribution. The mean, median and standard deviation of the Root Mean Square Errors are reported over the first $10$ estimation period.}
\end{table}

% table
\begin{table}[!htbp] \centering
\caption{New supervision tool validation, robustness to transformation - full Bayesian}\label{tab:validatetransform}

\subfloat[New supervision tool validation, robustness to scaling\label{validatescale}]{
\begin{tabular}{lrrr}
 & Pioneers & Linear & Median \\
2 & 1.00 & 2.62 & 1.34 \\
3 & 1.00 & 2.33 & 1.22 \\
4 & 1.00 & 1.48 & 1.14 \\
5 & 1.00 & 1.31 & 1.12 \\
6 & 1.00 & 1.24 & 1.11 \\
7 & 1.00 & 1.20 & 1.10 \\
8 & 1.00 & 1.18 & 1.10 \\
9 & 1.00 & 1.17 & 1.09 \\
\end{tabular}

}

\subfloat[New supervision tool validation, robustness to non-linear transformation\label{validatelog}]{
\begin{tabular}{lrrr}
 & Pioneers & Linear & Median \\
2 & 1.00 & 1.68 & 1.49 \\
3 & 1.00 & 1.16 & 1.05 \\
4 & 1.00 & 0.74 & 0.80 \\
5 & 1.00 & 0.60 & 0.73 \\
6 & 1.00 & 0.59 & 0.72 \\
7 & 1.00 & 0.61 & 0.72 \\
8 & 1.00 & 0.63 & 0.74 \\
9 & 1.00 & 0.65 & 0.74 \\
\end{tabular}

}

\subfloat[New supervision tool validation, relevance in a Gaussian context\label{validategaussian}]{
\begin{tabular}{lrrr}
 & Pioneers & Linear & Median \\
2 & 1.00 & 0.58 & 0.70 \\
3 & 1.00 & 0.74 & 0.88 \\
4 & 1.00 & 0.65 & 0.77 \\
5 & 1.00 & 0.73 & 0.87 \\
6 & 1.00 & 0.80 & 0.96 \\
7 & 1.00 & 0.87 & 1.04 \\
8 & 1.00 & 0.91 & 1.09 \\
9 & 1.00 & 0.95 & 1.14 \\
\end{tabular}

}

\Tabnote{The tail parameter is taken from a Pareto type one distribution with $\alpha=1.5$. Five non-cooperative Bayesian experts are modeled with independent observations from the loss distribution. $10^5$ Monte Carlo simulations are run. In Table \ref{validatescale}, estimates have been scaled by $100$. In table \ref{validatelog}, the estimates have been transformed with a logarithm $\log(c+\alpha)$ with $c=2$ a constant to avoid issues when $\hat{\alpha}$ are close to $0$. In table \ref{validategaussian}, the loss samples are taken from a standard normal law.}
\end{table}

% table
\begin{table}[!htbp] \centering
\caption{New supervision tool robustness checks - full Bayesian}\label{tab:robustnesschecks}
\begin{tabular}{rrrrr}
\toprule
$\alpha$ & experts & Linear RMSE & Median RMSE & Pioneers RMSE \\
\midrule
     0.5 &       2 &        3.48 &        3.48 &          1.00 \\
     0.5 &       5 &        3.53 &        1.51 &          1.00 \\
     0.5 &      20 &        3.08 &        1.21 &          1.00 \\
     1.1 &       2 &        3.29 &        3.29 &          1.00 \\
     1.1 &       5 &        3.51 &        1.54 &          1.00 \\
     1.1 &      20 &        3.25 &        1.26 &          1.00 \\
     2.0 &       2 &        3.19 &        3.19 &          1.00 \\
     2.0 &       5 &        3.70 &        1.56 &          1.00 \\
     2.0 &      20 &        3.73 &        1.30 &          1.00 \\
     3.0 &       2 &        3.03 &        3.03 &          1.00 \\
     3.0 &       5 &        3.57 &        1.41 &          1.00 \\
     3.0 &      20 &        4.52 &        1.34 &          1.00 \\
\bottomrule
\end{tabular}

\Tabnote{The tail parameter $\alpha$ and the number of Bayesian experts are varied. The last three columns report the average Root Mean Square Errors for the competing opinion pooling tools over the three initial estimation periods.}
\end{table}

% table
\begin{table}[!htbp] \centering
\caption{New supervision tool validation}\label{tab:relatedts}
\begin{center}
\begin{tabular}{llll}
 & x & y & z \\
Pioneers & 0.60 & 0.22 & 0.19 \\
Correlation & 0.46 & 0.48 & 0.06 \\
GC & 0.62 & 0.38 & 0.00 \\
\end{tabular}

\end{center}
\Tabnote{The time series are random sample of length $10$. $b=d=e=.9$ and $a=c=.1$. $10^5$ Monte Carlo simulations are run and the average of each weights are reported. The weights per method are expected to be ranked such that on average $w_x > w_y > w_z$.}
\end{table}

% table
\begin{table}[!htbp] \centering
\caption{New supervision tool validation}\label{tab:relatedtsinversed}
\begin{center}
\begin{tabular}{llll}
 & x & y & z \\
Pioneers & 0.31 & 0.21 & 0.48 \\
Correlation & 1.28 & 0.09 & -0.37 \\
GC & 0.50 & 0.44 & 0.06 \\
\end{tabular}

\end{center}
\Tabnote{The time series are random sample of length $10$. $b=d=e=.1$ and $a=c=.9$. $10^5$ Monte Carlo simulations are run and the average of each weights are reported. The weights per method are expected to be ranked such that on average $w_x > w_y > w_z$.}
\end{table}

	%appendix
	%\input{appendix_climate}

\clearpage
\newpage
\thispagestyle{empty}
\appendix
\renewcommand{\thesection}{\Roman{section}}%
\renewcommand{\thesubsection}{Appendix \Alph{subsection}}%
\renewcommand{\thesubsubsection}{\Alph{subsection}.\arabic{subsubsection}.}%

\setcounter{section}{0}
\setcounter{table}{0}
\setcounter{figure}{0}
\clearpage
\renewcommand{\thefigure}{A.\arabic{figure}}%

\section{Appendix}
\subsection{The Pioneer Detection Method and alternative novel approaches}\label{sec:alternativenovelapproaches}

\subsubsection{Pioneer Detection Method convergence properties}\label{sec:convergenceprop}

We show that the new supervision tool has some desired convergence properties. $m$ represents the number of experts and $t$ the observation periods which are also the observations count.

\emph{Property 1. When Bayesian experts' claims are independent and identically distributed (iid), as $m\to \infty$, the PDM converges to the mean of the experts' estimations.}

Writing $S$'s subjective opinion $o_t$, it is constructed with the PDM assigning a vector $w_t$ to the experts judgment $\hat{A}_t$, $o_t = \hat{A}_t w_t$. As each expert posterior of $\alpha$ follows a Gamma distribution with all the same shape $t$ and a rate parameter that tend to  $\frac{1}{\alpha}$ as $t\to\infty$, then as demonstrated in \citet{mathai1982storage}, an omniscient expert opinion will follow a Gamma distribution with $m t$ as a shape parameter, hence the mean of the experts' opinions. As $m\to\infty$, the Bayes omniscient expert opinion will converge to the true value of $\alpha$, \citet{de2003bayesian} find this Bayesian approach robust and having lower variance than Maximum Likelihood Estimation. Then as $m\to\infty$, it becomes impossible for any expert to lead the omniscient expert, which is also the average of all other experts, hence the PDM applies a weight of $1$ to the omniscient expert and $0$ on all experts, which is equivalent to assigning a weight $\frac{1}{m}$ to each expert's opinion. 

\emph{Property 2. When Bayesian experts' claims are iid, as $t\to \infty$, the PDM converges to the mean of the experts' estimations.}

For an expert, his posterior's rate parameter is an exponential autoregressive process as defined in \citet{gaver1980first}, with a unit root: $r_i^{t+1}=r_i^t+\epsilon_i^t$ with $\epsilon_i\sim \mbox{Exp}(\alpha), \forall i$. As such, $cov\left(r_i^{t+1}, r_i^{t} \right) = \frac{1}{\alpha^2} t$ hence as $t\to\infty$ the covariance between two experts estimates cannot be distinguished as leading any other and all experts will be treated equivalently by the PDM, hence $w_i^t \xrightarrow[t\to \infty]{} \frac{1}{m}$. 

\subsubsection{Weighing convergence with distances or angles}\label{sec:speed}

\begin{figure}[H]
\caption{Alternative to the orientation change dummy to identify Pioneers}\label{fig:step3alternative}

\begin{tikzpicture}[scale=0.7] \begin{axis}[
    xlabel={Time steps},
    xtick={-2,-1,0},
    ylabel={Risk parameter},
    ]
    \addplot+[blue] coordinates {
        (-2,17) (-1,14) (0,9)
    };
    \addplot+[red] coordinates {
    (-2,10) (-1,7.5) (0,5) 
    };
    \legend{$-i$,$i$}
    \addplot+[red, dashed] coordinates {
        (-1,7.5) (0,7.5) 
    };
        \addplot+[blue, dashed] coordinates {
         (-1,14) (0,14)        
    };
    \coordinate (A) at  (0,14) ;
    \coordinate (B) at (-1,14);
    \coordinate (C) at (0,9);
    \pic [draw, ->, "$\theta_{-i}^t$", angle radius=1.7cm, angle eccentricity=1.5] 
    {angle = C--B--A};
    \coordinate (D) at  (0,7.5) ;
    \coordinate (E) at (-1,7.5);
    \coordinate (F) at (0,5) ;
    \pic [draw, ->, "$\theta_{i}^t$", angle radius=1.7cm, angle eccentricity=1.5] 
    {angle = F--E--D};
\end{axis}
\end{tikzpicture}
\qquad
\begin{tikzpicture}[scale=0.7] \begin{axis}[
    xlabel={Time steps},
    xtick={-2,-1,0},
    ylabel={Risk parameter},
    ]
    \addplot+[blue] coordinates {
        (-2,17) (-1,14) (0,9)
    };
    \addplot+[red] coordinates {
    (-2,10) (-1,7.5) (0,5) 
    };
    \legend{$-i$,$i$}
    \addplot+[red, dashed] coordinates {
        (-1,7.5) (0,7.5) 
    };
        \addplot+[blue, dashed] coordinates {
         (-1,14) (0,14)        
    };
    \coordinate (A) at  (0,14) ;
    \coordinate (B) at (-1,14);
    \coordinate (C) at (0,9);
    \draw[dashed,<->] (A) -- (C);
    \node (A) at  (0,14){};
\node(B) at (-1,14){};
\node (C) at (0,9){};
    \node (D) at  (0,7.5) {};
    \node (E) at (-1,7.5){};
    \node (F) at (0,5) {};
%\path[-angle 90,font=\scriptsize]
\path[->] 
(A)   edge [left]  node {$\Delta_{-i}^t$} (C)
(D)   edge [left]  node {$\Delta_i^t$} (F);
\end{axis}
\end{tikzpicture}

\begin{minipage}{.7\linewidth}
\footnotesize
\emph{Notes:} $i$ represents the expert of interest and $-i$ the average estimate of his competitors ($i$ excluded). The weight in the left panel is $\frac{\vert \theta^t_{-i}\vert  }{\vert \theta^t_{-i}\vert +\vert \theta^t_{i}\vert } $. The weight in the right panel is $\frac{\vert \Delta^t_{-i}\vert  }{\vert \Delta^t_{-i}\vert +\vert \Delta^t_{i}\vert } $
\end{minipage}
\end{figure}
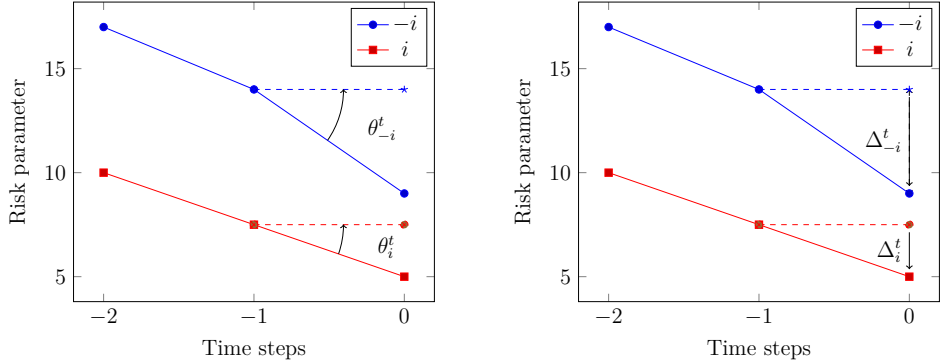

In the last two steps of the PDM, the weights can be defined with angles or distances as illustrated in the right panel Figure \ref{fig:step3alternative}. Angles are the preferred approach because they allow the supervisor to take into account the speed of convergence between time series. If $\theta$ is the angle between vector $\vec{u}$ and $\vec{v}$, it can be computed as
\begin{equation}\label{eq:anglevectors}
\begin{split}
\theta &=\cos^{-1}\left( \frac{u_x v_x + u_y v_y}{\sqrt{u_x^2 + u_y^2}\sqrt{v_x^2+v_y^2}} \right)\\
 & = \cos^{-1}\left( \frac{s^2+ u_y v_y}{\sqrt{s^2 + u_y^2}\sqrt{s^2+v_y^2}} \right)
\end{split}
\end{equation}
the distance relevant to the measure are taken from the y-axis $u_y$ and $v_y$ and the weight is computed as $\frac{\vert v_y\vert}{\vert u_y\vert + \vert v_y\vert}$ which do not include the x-axis $u_x=v_x=s$ the step between two observation points. This weighting with distance is found to be non-robust, as shown in Table \ref{tab:validatemeasure}.

Alternative inter-temporal pioneers detection methods could also be implemented with more traditional time series methods. I list the five candidate methods and demonstrate how they boil down to implementing the Granger Causality, lagged-correlation and probabilistic combinations methods as alternatives to the PDM.

\subsubsection{Granger Causality}

The core principle of the PDM entails assigning indirect votes based on expert estimates. While experts are non-cooperative and do not exert influence on one another, this approach exhibits certain similarities with the identification of Granger Causality, which tests whether a given time series is beneficial in predicting another. It examines whether an expert's opinion change appears to precede a similar opinion change from his competitors. I implement the test introduced by \citet{granger1969investigating}. \citet{toda1995statistical} take into account potential integration and cointegration between time series. \citet{hasbrouck1995one} also introduces cointegration to determine the information share of each random variable.
In the context of climate change and small time series (maximum lag of $2$), the integration or cointegration orders cannot be tested robustly. Therefore, I assume that the time series are locally stationary and apply a Granger Causality test with a lag of one. %The limited history available for $D$ to make a decision renders this method a weak candidate.

\subsubsection{Lagged Correlation}

An alternative approach involves measuring correlations between lagged estimates from each expert and the estimates from his competitors. This can be measured using the Pearson coefficient \citep{pearson1895}, as in \citet{sakurai2005braid} and \citet{forbes2002no} for financial applications.

\subsubsection{Probabilistic combinations}

When estimates are not points but probabilistic distributions, on top of the above methods, two additional methods are considered. The first is the Bayesian Model Averaging (BMA, \citet{draper1995assessment}) which presents three challenges \citep{wang2022forecast}, a blocking one for my approach is to elicit a prior which comes back to doing a combination choice. Quantile combinations \citep{vincent1912functions} present the advantage to keep location-scale family after the transformation. I apply the vincentization as a candidate in a probabilistic set up, results are available from the author.

\subsubsection{Multivariate Linear Regressions}

A related approach to the Granger causality test involves using multivariate linear regressions of each expert's estimates on the average estimate from his competitors, as in \citet{yi2000online}. If significant, the coefficients can be considered as voting weights for each expert. This approach with limited history returns to searching for correlation and Granger causality between time series.

\subsubsection{Information Transfer}

I also explore the information transfer literature and measures \citep{schreiber2000measuring}. The idea is to measure whether the time series act as if there were transfer entropy, that is, information transport from one series to another. For financial applications with non-Gaussian random variables, this approach requires discretizing continuous time series with bins. \citet{dimpfl2014impact} limit their analysis to three bins and divide the return data along the 5\% and 95\% quantiles, as they ``assume that extreme (tail) events are more informative than the median observation."

The information transfer method tests whether an expert's opinion change appears to be informative to his competitors. With non-cooperative experts, no information is exchanged, and therefore, if an apparent information transfer is detected, it is as if one expert learnt from the Data Generating Process (DGP) before his competitors, and then the DGP informs the competitors, which is similar to the expert directly informing his competitors if the expert learns faster. \citet{barnett2009granger} demonstrate that this method is similar to Granger causality if the random variables are Gaussian. When experts are Bayesian, the posteriors are the random variables of interest and can be approximated as Gaussian, hence I implement the Granger causality method.

\subsection{Pioneer Detection Method and linearly related time series}\label{sec:linearts}

I test the PDM capacity to identify linearly related time series. I define three time series as the experts $x$, $y$ and $z$ in the spirit of \citep{Granger:1974aa}:
\begin{equation}
\begin{split}
x_t &= x_{t-1} + \epsilon_t\\
y_t &= a y_{t-1} + b x_{t-1} + \nu_t\\
z_t &= c z_{t-1} + d y_{t-1} + e x_{t-1} + \xi_t\\
\end{split}
\end{equation}
with $\epsilon$, $\nu$ and $\xi$ white noises, I set $x_0=y_0=z_0=0$. The results are sensitive to the coefficients, I set auto-regressive coefficients significantly lower than cross linear relationships so the cross-relationships can be captured by the method.
$x$ is the main pioneer of the group as the innovations $x$ faces are passed on to other time series as $x$ has a unit root. On the contrary, $y$ has an auto-regressive coefficient below unity and his innovations are transient. The main aim of this test is to confirm that all methods can identify the pioneership of $x$ and measure to which extent the effect of $y$ on $z$ can be identified.
Table \ref{tab:relatedts} reports the average weight each of the three novel Pioneer Detection Methods assigns to each time series. Table \ref{tab:relatedtsinversed} reports the detection capacity when the coefficient has a lesser dampening effect on noise.

\end{document}